\documentclass[12pt]{article}

\usepackage[margin=1in]{geometry}
\usepackage[round,sort]{natbib}
\usepackage{subcaption}
\usepackage{graphicx}
\usepackage[hidelinks]{hyperref}
\usepackage{xurl}
\usepackage{authblk}
\graphicspath{{./plots/}}

\title{Auditing the Use of Language Models\\to Guide Hiring Decisions\thanks{%
  We are grateful to Avi Bagchi and Marissa Gerchick for research assistance.
}}
\author[1]{Johann D. Gaebler}
\affil[1]{Department of Statistics, Harvard University}

\author[2]{Sharad Goel}
\affil[2]{Kennedy School of Government, Harvard University}

\author[3]{Aziz Huq}
\affil[3]{University of Chicago Law School}

\author[4]{Prasanna Tambe}
\affil[4]{The Wharton School, University of Pennsylvania}

\date{}

\begin{document}

\maketitle

\begin{abstract}
  \noindent
  Regulatory efforts to protect against algorithmic bias have taken on increased
  urgency with rapid advances in large language models (LLMs), which are machine
  learning models that can achieve performance rivaling human experts on a wide
  array of tasks. A key theme of these initiatives is algorithmic ``auditing,''
  but current regulations---as well as the scientific literature---provide
  little guidance on how to conduct these assessments. Here we propose and
  investigate one approach for auditing algorithms: correspondence experiments,
  a widely applied tool for detecting bias in human judgements. In the
  employment context, correspondence experiments aim to measure the extent to
  which race and gender impact decisions by experimentally manipulating elements
  of submitted application materials that suggest an applicant’s demographic
  traits, such as their listed name. We apply this method to audit candidate
  assessments produced by several state-of-the-art LLMs, using a novel corpus of
  applications to K-12 teaching positions in a large public school district. We
  find evidence of moderate race and gender disparities, a pattern largely
  robust to varying the types of application material input to the models, as
  well as the framing of the task to the LLMs. We conclude by discussing some
  important limitations of correspondence experiments for auditing algorithms.
\end{abstract}

\thispagestyle{empty}

\newpage

\section*{Introduction}

AI-based systems have the potential to assist employers with many aspects of
human resources (HR) management, from benefits administration to coaching and
development to its most common HR use case, applicant screening. The global HR
technology market based on predictive models was already rapidly growing prior
to 2022, but attention to AI tools received a dramatic boost with the advent of
large language models (LLMs), which are models that are highly adept at
understanding, summarizing, and evaluating text data. Given the primacy of text
data in the job application process, an emerging HR use case for modern LLMs is
to ingest entire application dossiers---including resumes, essays, and
transcripts captured from interviews---and output seemingly cogent assessments
of candidates' qualifications.

As hiring use cases proliferate, however, employers and policymakers are racing
to establish guidelines around whether the algorithmic evaluation of candidates
comports with employment discrimination law, and how to audit commonly deployed
AI tools to ensure they are not discriminatory. The ethical and legal
implications of using predictive tools in HR has motivated a body of academic
work~\citep{tambe2019artificial, raghavan2020mitigating}. Policymakers have
matched the attention of firms and researchers, introducing a wave of
legislation governing high-stakes algorithmic decision making, and hiring in
particular (e.g., New York LL 144 or Illinois 820 ILCS 42). Nevertheless, there
are still few tools for identifying potentially biased decision making in LLMs,
in part because their inner workings are opaque.

In contrast to traditional supervised machine-learning algorithms, LLMs can
generate hiring recommendations even absent human-generated candidate ratings.
Supervised algorithms start with training datasets that contain both application
materials and ratings for a subset of applicants. From the training data, these
algorithms learn patterns in the relationship between inputs and ratings to
predict human assessments of unrated candidates. LLMs, on the other hand, are
``pre-trained'' algorithms that rely on an immense corpus of generalized
training data to produce results with only high-level instructions describing a
task. In this way, LLMs are closer to human evaluators, who might produce
candidate ratings based on an intuitive understanding of how professional
experiences and responses to interview questions relate to competency and fit
for a position. This generality of LLMs underpins their potential for HR and
beyond. But it also raises new concerns that LLMs might produce discriminatory
or distorted responses in ways that are hard to predict.

Even without these added complexities, there is no scientific consensus on how
best to audit algorithms for bias. Researchers have proposed a wide variety of
algorithmic fairness metrics~\citep{%
  dwork2012fairness, hardt2016equality, kleinberg2016inherent,
  corbett2017algorithmic, chouldechova2017fair, kilbertus2017avoiding,
  kusner2017counterfactual, zafar2017fairness, loftus2018causal, nabi2018fair,
  chiappa2019path, wang2019equal, chouldechova2020snapshot,
  coston2020counterfactual, imai2020principal,  berk2021fairness,
  corbett2023measure, raghavan2023should%
}.
Here, instead, we study the potential application of so-called ``correspondence
experiments''---popular in the behavioral sciences for identifying
discrimination in human decisions---to audit LLMs for potential race and gender
bias in high-stakes decision settings like HR.\@ Correspondence experiments
(also known as ``audit studies'') proceed from the assumption that two otherwise
identical individuals from different demographic groups should receive similar
decisions, and that divergent treatment is evidence of improper discrimination.
To operationalize this idea, researchers typically focus on settings where
decision makers interact with individuals exclusively through written documents
(e.g., an initial screening of job applicants), and then experimentally
manipulate elements of those materials that suggest an individual's race and
gender, such as their listed name or pronouns~\citep{gaddis2019understanding}.
For at least fifty years, social scientists and government agencies have
employed correspondence experiments to study discrimination in
hiring~\citep[e.g.,][]{bertrand2004emily},
housing~\citep[e.g.][]{wienk1979measuring}, prosecutorial charging
decisions~\citep[e.g.,][]{chohlas2021blind}, and other
domains~\citep{ayres2015race, gaddis2020searching, lyons2019race}. More
recently, correspondence experiments have been proposed to similarly identify
algorithmic bias~\citep{haim2024s, tamkin2023evaluating, veldanda2023emily}.

We adapt this methodological approach to audit LLM-based assessments of job
applicants. We start with a novel corpus of real job applications to K-12
teaching positions at a large Texas public school district, which includes
applicant resumes as well as video responses to interview questions. Based on
these materials, we elicit hiring recommendations from several state-of-the-art
open-source and proprietary LLMs. By experimentally manipulating names and
pronouns in the application materials, we find evidence of moderate race and
gender effects in the algorithmic ratings. These results illustrate the
potential of correspondence experiments to serve as a concrete assessment
strategy under existing algorithmic audit mandates. However, we also discuss
some key conceptual and technical limitations of this approach for conducting
algorithmic audits.

\subsection*{Regulatory Landscape}

Since 2021, local, national, and supranational regulators have proposed and
enacted private and public AI regulation. Auditing demands occur in both
American and European law. What it means to ``audit'' an AI, however, often
remains vague. Three examples are noteworthy.

First, the most developed audit mandate arises under a New York City ordinance,
Local Law 144 (``LL144''), effective July 2023. LL144 requires a ``bias audit''
when employers use ``any computational process, derived from machine learning,
statistical modeling, data analytics, or artificial intelligence'' to classify
or recommend persons for employment. Bias audits by independent third-parties
must calculate and publicly report an ``impact ratio,'' i.e., the rate at which
individuals in a race or gender category are positively selected relative to the
selection rate of the most positively treated category. LL144 imposes no legal
obligations when a disparity is identified. Employers also have unfettered
discretion to determine whether they are covered by the measure. In the first
six months after LL144's entry into force, only nineteen audits linked to the
law were published~\citep{groves2024auditing}.
\nocite{ll144}

Second, on October 30, 2023, President Biden issued an executive order imposing
mandates on federal agencies' use of AI and directing regulatory efforts by
agencies with authority over private AI uses. The executive order twice mentions
audits as tools to advance fair public decision-making and to ensure AI safety.
First, the Agriculture Secretary is required to issue ``guidance'' on AI use in
public benefits programs to ``enable auditing and, if necessary, remediation of
the logic used to arrive at an individual decision.'' Second, the Secretaries of
Commerce, Energy, and Homeland Security, must promulgate ``guidelines and best
practices \dots\ evaluating and auditing [private] AI capabilities, with a focus
on capabilities through which AI could cause harm, such as in the areas of
cybersecurity and biosecurity.'' As of April 2024, no guidelines had been issued
under either provision. Follow-on regulations from the Secretaries, moreover,
may offer only general indication about how to implement audits.
\nocite{whitehouse2023}

Finally, European law includes different audit requirements. The Digital
Services Act (``DSA''), which entered into force in February 2024, requires
“very large online platforms” to conduct ``audits'' for ``compliance'' with the
DSA's requirements. In March 2024, the European Parliament adopted a more
general AI Act. The AI Act imposes a variety of requirements that potentially
involve audits, most notably compelling users of “high risk” systems, including
those that ``profile'' individuals or make decisions related to health and
personal economic situations, to create ``quality management systems'' that
appear to be a kind of audit.
\nocite{dsa2024, aiact2024}

\section*{Empirical Analysis and Results}

\subsection*{LLMs for Candidate Evaluation}

We illustrate both the use of LLMs for hiring and the use of correspondence
experiments to audit these hiring algorithms. We start with a novel corpus of
1,373 applications to K-12 teaching positions at a large public school district
in Texas, which we collected via a public records request. Application materials
include the applicant's resume and self-recorded video responses to questions
asking about previous teaching experience, teaching style, hypothetical
classroom situations, and other job-related subjects. From these videos, we
automatically generate written transcripts of each candidate's responses using
speech-to-text software. We restrict our analysis to the 801 applicants who both
provided a resume and underwent a video interview. These applicants comprise a
diverse pool: 67\% of applicants are women, 45\% are Black, 10\% are Hispanic,
and 39\% are White.

The school district to which the candidates applied does not, to our knowledge,
use algorithms to evaluate applicants. However, to demonstrate both the
potential use of LLMs in this context, as well as potential algorithmic auditing
procedures, we implement a simple automated candidate evaluation pipeline. In
particular, for each applicant, we input to an LLM: (1) a description of the
requirements for the teaching position based on a job posting from the school
district; (2) the applicant's resume; (3) a written transcript of the
applicant's self-recorded responses to interview questions; (4) a request for
the model to summarize the candidate's qualifications in prose; and (5) a
request for the model to provide numerical evaluations, on a scale from 1 to 5,
of the candidate's experience, professionalism, and fit, as well as the model's
overall hiring recommendation, again ranging from 1 (``definitely do not hire'')
to 5 (``definitely hire''). We restrict our primary statistical analysis to the
overall numerical hiring recommendation produced by the model, but we elicit the
free-text summary and additional numerical scores to encourage higher-quality
results, a strategy often referred to as ``chain-of-thought'' prompting.

We apply this mode of algorithmic assessment to our 801 applications using
OpenAI's \texttt{GPT-3.5} LLM for our initial analysis. It is difficult to
determine the extent to which the model's ratings reflect a candidate's
qualifications, and such an assessment is not the focus of our analysis here.
But, based on informal inspection, the ratings have face validity, with the
highest rated candidates generally having more experience and giving more
polished responses to interview questions than those receiving lower ratings. It
thus seems reasonable to assume that a pipeline like the one we implement here
will soon be used by employers to screen applicants---if that is not already
happening.

\subsection*{Assessing Adverse Impacts}

Presented with such model-generated candidate assessments, the first question
one might ask in any algorithmic audit is whether the model rates candidates
similarly across demographic groups. Following guidance in New York City's
LL144, as well as state and federal disparate impact statutes, we start by
computing the adverse impact ratio: the rate at which individuals in one race or
gender category are ``positively selected'' (e.g., moved to the next interview
stage) relative to those in another. The Equal Employment Opportunity
Commission's ``four-fifths'' rule flags an adverse impact ratio of 80\% or lower
as particularly concerning~\citep{tobia2017disparate}. For simplicity, we
consider only the summary ``hiring'' rating produced by the LLM, and convert the
numerical score to a binary assessment by thresholding the rating at either 3,
4, or 5. For example, at a threshold of 4, we consider as ``positively
selected'' those candidates who score 4 or above. Overall, 23\% of candidates
received a 5, 57\% received a 4, and 25\% received a 3, and 5\% received a 2; no
candidate received a 1.

\begin{figure}[t]
  \begin{center}
    \includegraphics{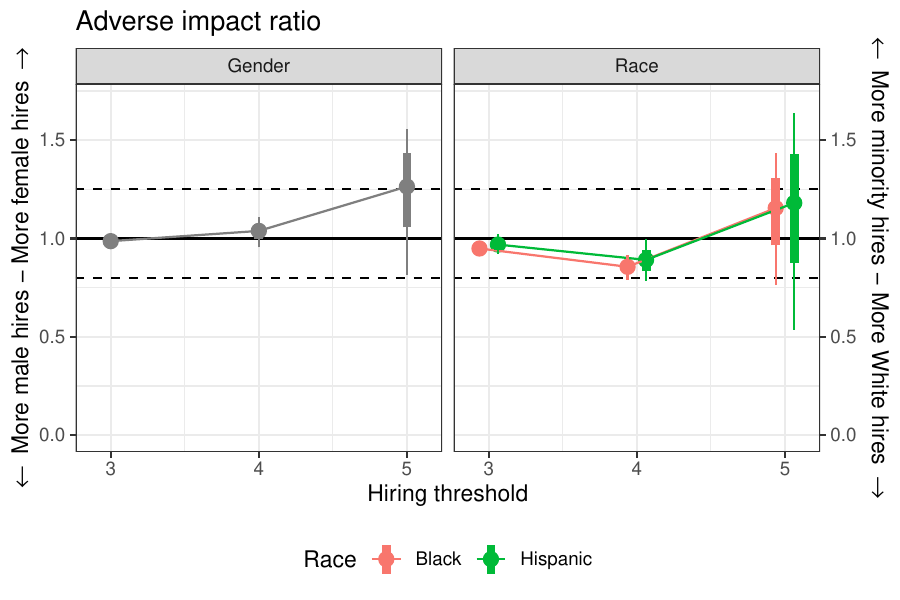}
  \end{center}
  \caption{\emph{%
    Adverse impact ratios for LLM hiring recommendations at different hiring
    thresholds, with pivotal 95\% bootstrapped confidence intervals. At the
    lowest threshold, we observe near parity across both race and gender;
    however, at higher thresholds, we find some evidence of disparities in
    hiring rates across demographic groups, though the estimates are imprecise.
  }}%
\label{fig:adverse-unblinded}
\end{figure}

The results of this adverse impact analysis are shown in
Figure~\ref{fig:adverse-unblinded}. At the highest threshold (i.e., comparing
the proportion of applicants across groups who received a 5), we find that
female applicants received positive assessments more often than men, and Black
and Hispanic applicants received positive assessments more often than White
applicants, though in all these cases the estimates can only be imprecisely
estimated with our available data. At a threshold of 4, the pattern flips for
race---i.e., White applicants received a positive assessment more often---and we
see near-parity for gender. And at a threshold of 3, we find near parity across
both gender and race groups. 

This simple analysis suggests that the model might be favoring certain
demographic groups. Though, without further evidence, we cannot definitively say
whether these disparities are due to algorithmic bias or group-specific
differences in the applicant pool. Women and racial minorities in our applicant
pool might, in actuality, be more qualified for these positions, potentially
explaining the higher model ratings they receive.

\subsection*{Correspondence Experiments}

To tease apart these two explanations---algorithmic bias versus differences in
candidate qualifications---we conduct a correspondence experiment. We start by
manipulating the real application materials to create synthetic dossiers that
differ from the real applications only in details that strongly signal an
applicant's race or gender. More specifically, for each real applicant, we
generate eight synthetic applications, corresponding to a particular race
(Asian, Black, Hispanic, or White) and gender (female or male). We then replace
the applicant's actual name throughout the application materials with one that
strongly signals membership in that group. We similarly change any mention of
the applicant's pronouns in the materials to match the assigned group, as well
as other indicia of race or gender. (See Appendix for a detailed description of
how we generate these synthetic applications.)

To ensure that our manipulation worked as intended, we presented the synthetic
application materials to the LLMs we considered, this time instructing the
models to report the race and gender of the synthetic applicants. We find
generally high agreement between our intended race and gender and the model's
“perception” of these attributes, in most cases exceeding 90\%, but with the
precise level of agreement varying from case to case; see
Figures~\ref{fig:manip-check-gender} and~\ref{fig:manip-check-race} in the
Appendix. This level of agreement is comparable to the effectiveness of
manipulating human perception of race by altering names in
resumes~\citep[cf.][]{bertrand2004emily}.

To test the extent to which the model's evaluations are influenced by race and
gender---as signaled through listed names and other manipulated elements---we
again elicit hiring recommendations, this time of the synthetic candidates. We
audit eight LLMs: OpenAI's \texttt{GPT-3.5} and \texttt{GPT-4}
models~\citep{brown2020language, achiam2023gpt}, Mistral's \texttt{Mistral~7B}
and \texttt{Mixtral~8x7B} models~\citep{jiang2023mistral, jiang2024mixtral}, and
Anthropic's \texttt{Claude~Instant}, \texttt{Claude~2}, \texttt{Claude~3~Haiku},
and \texttt{Claude~3~Sonnet} models~\citep{anthropic2023, anthropic2024}.
OpenAI's and Anthropic's models are considered to be the best currently
available, and Mistral's models are popular open-source competitors.

\begin{figure}[t]
  \begin{center}
    \includegraphics{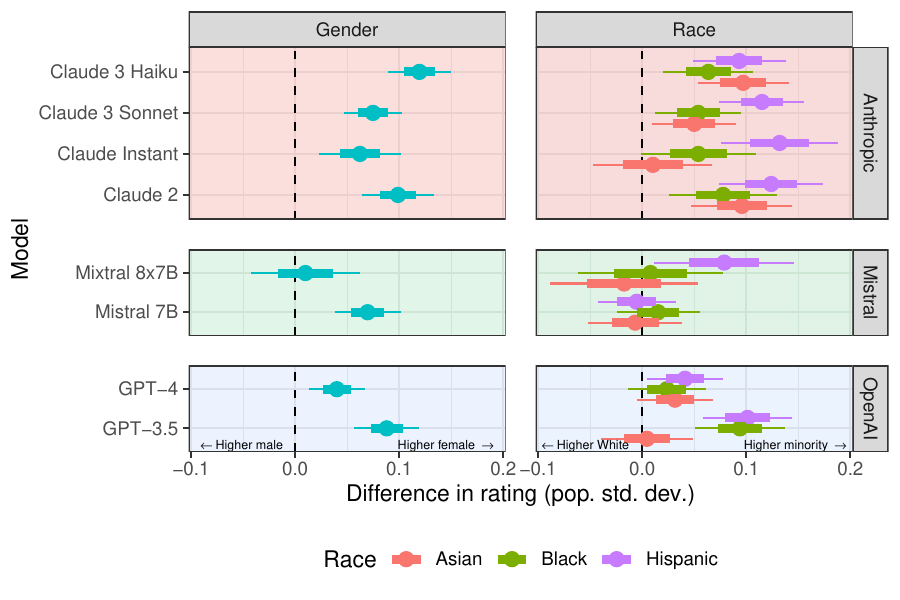}
  \end{center}
  \caption{\emph{%
    Differences in mean model scores across LLMs between synthetic applicants of
    different races and genders, reported in estimated population standard
    deviations, with 70\% and 95\% confidence intervals clustered by the real
    application dossier used to generate the synthetic application. Positive
    values indicate that the model rates female or racial minority applicants
    higher than male or White applicants on average.
  }}%
\label{fig:models}
\end{figure}

The results of these correspondence experiments are shown in
Figure~\ref{fig:models}. For each model, we plot the standardized estimated
difference in mean model score between a given gender or race group and the
reference group, where the reference group for gender is men and the reference
group for race is White people. (See the appendix for further methodological
details.) Across models, we find that the LLMs rate the synthetic female
candidates moderately higher than the synthetic male candidates. Turning to
race, we find that the models generally rate synthetic Black, Hispanic, and
Asian candidates moderately higher than synthetic White candidates, though we
find more variation across models, with Mistral's models exhibiting smaller
disparities. Our correspondence experiments thus suggest that race and gender
influence our algorithmic candidate assessments, at least to some degree.

\subsection*{Sensitivity to Prompt Variation and Context}

LLMs can be sensitive to the exact way in which responses are elicited. To
assess the robustness of our findings, we repeat our analysis with several
variants of the prompt. First, we generate several distinct but substantively
similar prompts via a separate model that we prompted to produce these
variations. Second, we generated a variant that omitted the intermediate
candidate summarization step in our primary prompt. Third, we added to our
prompt an explicit statement instructing the model to follow anti-discrimination
protections outlined in the EEOC guidelines. For simplicity, we re-ran these
robustness tests only on one model, OpenAI's GPT-3.5, a model that exhibited
approximately average disparities in our primary analysis above. We found the
general patterns above persisted across all these variants; see
Figures~\ref{fig:variant} and~\ref{fig:framing} in the Appendix.

In some cases, employers, during initial screenings, may have access only to
candidate resumes and not interviews. To gauge potential model disparities in
that setting, we re-ran our primary analysis while inputting only an applicant's
resume into the model. As above, we conducted this analysis on GPT-3.5, finding
again that women and racial minorities received moderately higher scores than
men and White applicants, respectively.

Many LLMs can access virtually encyclopedic knowledge, including many details
about the specific school district to which the applicants applied, and
incorporate this into their evaluations. The Texas school district to which our
candidates applied is especially racially diverse, potentially interacting with
candidate demographics in unexpected ways. To test for the possibility that the
model's evaluations reflect student demographics or other details specific to
the school district, we replaced all mentions of the district (and the city and
state) with the name of a predominately White school district in West Virginia.
As above, we ran this analysis on GPT-3.5, and again found disparities mirroring
our primary results.

\section*{Discussion}

Our results demonstrate the potential of correspondence experiments to identify
algorithmic gender and race bias, offering policymakers a tool for auditing
algorithms. The substantive patterns we observe appear robust to several
modifications of our primary study, including variations on the model
instructions we provide and the specific application materials we input.

Caution, however, is warranted when interpreting these results. Most
importantly, our findings may not generalize to other contexts where LLMs may be
applied. Indeed, while some recent studies report disparities similar to those
we find here~\citep{tamkin2023evaluating}, others report disparities that go in
the opposite direction~\citep{haim2024s, veldanda2023emily}. Such contrasting
results are not surprising given the complex and often inscrutable ways in which
LLMs are trained, including a final ``alignment'' phase where developers
fine-tune models to produce ``desirable'' outputs. This step is key to avoiding
the often overtly discriminatory results of unaligned models, but may also leave
traces of bias in ways that are hard to predict. Our results thus illustrate the
need to audit LLMs for each specific application, with an understanding that
conclusions may be sensitive to both the task and the pool of individuals
evaluated. 

Anti-discrimination norms commonly applied by employment law seek to ensure that
similarly \emph{qualified} applicants from different gender or race groups are
treated similarly. But because a candidate's ``qualifications'' are inherently
hard to measure, correspondence experiments instead probe whether assessments
are comparable across demographic groups, \emph{all else being equal}. This is
typically a conservative measure of disparate impact. Imagine that an LLM,
before producing its hiring recommendation, removes an applicant’s name,
pronouns, and other clear indicia of their demographic group from the input
materials---an approach sometimes called ``fairness through
unawareness''~\citep{dwork2012fairness}. A correspondence experiment that
manipulated applicant names would then (correctly) conclude that this
hypothetical algorithm is not impacted by gender or race---as signaled by listed
name. But that conclusion would provide limited assurances, as an applicant’s
demographic attributes might seep through and impact algorithmic assessments
through other channels to which the model had not been blinded.

In our own pool of applicants, we find that passing to the model “blinded”
application materials---with names and pronouns removed---has little effect on
the adverse impacts shown in Figure 1. However, hiding applicant race and gender
from the models is challenging---LLMs can accurately infer a candidate’s
demographics even from these blinded materials (see Appendix for details).
Consequently, even an algorithm designed to ignore clearly problematic
information---like an applicant’s name---could unjustly penalize candidates
based on their tacitly inferred group membership.

The call to audit algorithms is likely only to increase as LLMs become more
capable and widespread. Correspondence experiments, despite their important
limitations, are one method to audit algorithms for race and gender bias. We
hope our work aids current regulatory efforts to ensure these promising new
models yield equitable outcomes.

\bibliographystyle{plainnat}
\bibliography{refs}

\newpage
\appendix
\setcounter{figure}{0}
\renewcommand{\thefigure}{A\arabic{figure}}

\section{Data Redaction and Processing}

To facilitate the use of LLMs, we convert the application materials entirely to
text. Resumes were provided in a variety of formats, which we converted to PDFs,
and then hand-redacted to remove addresses, emails, phone numbers, other
individuals' names, and other references to personal information. We then
extracted the resulting resume contents using optical character recognition with
Amazon Textract. We also transcribed the video responses with automated speech
recognition using Amazon Transcribe. Finally, we manually code applicant race
and gender using interview videos, since self-reported race and gender are not
available. (We note that we only use manually coded demographic variables in the
adverse impact analysis.)

To manipulate how the applicant's demographic identity is presented to the
model, we mask various indicators of the applicants’ races and genders through a
combination of manual and automatic redaction in both the resumes and interview
transcripts. In particular, we remove applicants’ names, colleges (which might
signal race or gender if the applicant attended an HBCU or women’s college),
college locations, titles (e.g., ``Mr.'' or ``Mrs.''), and third-person pronouns
(e.g., ``she,'' ``her,'' or ``hers''). In experiments, we replace these elements
of the interviews to generate synthetic applications in which the applicant's
name and other elements are chosen to signal membership in a particular group,
as detailed below.

We also attempt to remove other information from the application materials that
may contradict the synthetic elements of the application. For instance, we also
redact information like whether the applicant has a husband or wife, is a mother
or father, and explicit references to their race- or ethnic-background or
appearance, replacing these with fixed placeholders. Similarly, we redact the
location of jobs that the applicant held during college, which might contradict
information we provide about where the synthetic applicant attended college.

\section{Synthetic Application Generation}

To generate a synthetic application dossier from a real application dossier, we
manipulate the applicant’s name, college, title, and third-person pronouns
appearing in the applicant’s resume and transcribed interview responses to
reflect a specific race (Asian, Black, Hispanic, White) and gender (female,
male).

\begin{figure}[t]
  \begin{center}
    \includegraphics{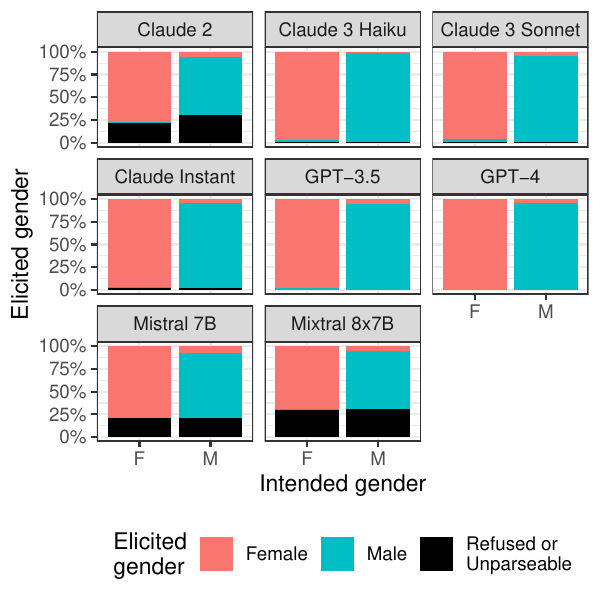}
  \end{center}
  \caption{\emph{%
    Agreement between the model’s ``perception'' of a synthetic applicant's
    gender and the gender we intended to associate with the synthetic applicant.
  }}%
\label{fig:manip-check-gender}
\end{figure}

The college we list the synthetic applicant as having attendedd is chosen
uniformly at random from the following list, without regard to race or gender:
\begin{itemize}
  \item The University of Houston in Houston, Texas,
  \item The University of Texas at Arlington in Arlington, Texas,
  \item The University of North Texas in Denton, Texas.
\end{itemize}
These universities were chosen according to the following criteria: (1) they are
relatively close in ranking among Texas universities, according to the US News
\& World Report, both to each other as well as to colleges and universities
attended by actual applicants to the school district; (2) their student bodies
are large and comparatively diverse, both racially and with respect to gender.

The title and pronouns we use in a synthetic application are chosen to match the
synthetic applicant’s gender. To present the applicant as female, we ensure that
in the resume and the transcripts of the interview they are referred to using
``she,'' ``her,'' or ``hers'' as appropriate, and as ``Ms.'' when a title is
used. (The transcripts record only the applicant; however, many applicants refer
to themselves in the third person when quoting students or colleagues.) To
present the applicant as male, we ensure that the pronouns ``he,'' ``him,'' and
``his'' and the title ``Mr.'' are used as appropriate.

Finally, we signal the applicant’s race and gender through a random choice of
one of twenty race- and gender-specific first- and last-name pairs. These pairs
are chosen to strongly signal the race and gender of the applicant. To find
names with this property, we draw real names from the North Carolina vote file,
which is publicly available.\footnote{%
  \url{https://www.ncsbe.gov/results-data/voter-registration-data}
}
Specifically, we choose a random sample of 100,000 first and last names, which
we then embed into 256 dimensions (512 dimensions total) using OpenAI’s
\texttt{text-embedding-small} text embedding model. Then, for each of the eight
possible choices of race and gender, we train a penalized logistic regression
model to predict the probability that someone has the chosen race and gender
using the embeddings. Then, we rank the names according to this probability,
discarding duplicate first names. Finally, we choose the top twenty name pairs
for each chosen race and gender.

\section*{Additional Results}

\subsection*{Manipulation Check}

\begin{figure}[t]
  \begin{center}
    \includegraphics{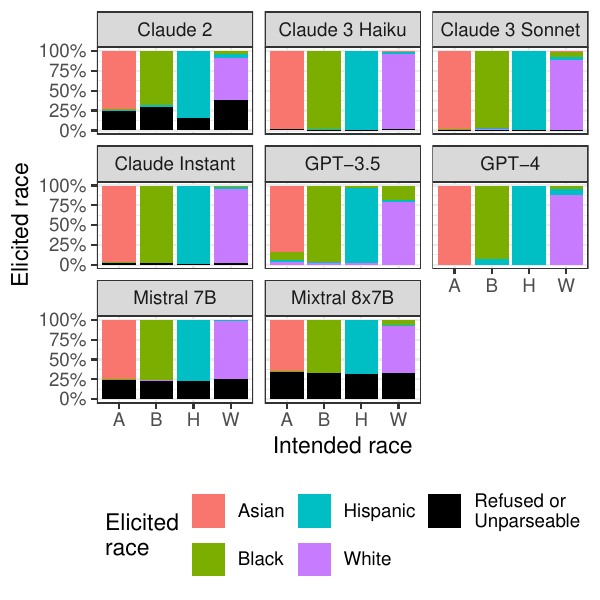}
  \end{center}
  \caption{\emph{%
    Agreement between the model’s ``perception'' of a synthetic applicant's race
    and the race we intended to associate with the synthetic applicant.
  }}%
\label{fig:manip-check-race}
\end{figure}

To confirm that our manipulation actually affects the model’s perception of race
and gender, we present the model with the manipulated application materials and
elicit the applicant’s race and gender only, rather than an evaluation of their
qualification for a teaching position. We then parse these responses using
\texttt{GPT-3.5} to obtain a structured representation of the model’s response.
The results are shown in Figures~\ref{fig:manip-check-gender}
and~\ref{fig:manip-check-race}.

As can be seen, the manipulation is successful most of the time. The exceptions
come from some of the less powerful models---viz., \texttt{Claude~2},
\texttt{Mistral~7B}, and \texttt{Mixtral~8x7B}---giving responses to the prompt
that do not contain the applicant's race or gender at all, either because the
response is nonsensical, or because the model refuses to answer the prompt as
directed. In cases where race and gender are provided, however, they closely
match the race and gender that we intend to ascribe to the synthetic applicant.

\subsection*{Prompt Variations}

\begin{figure}[t]
  \begin{center}
    \includegraphics{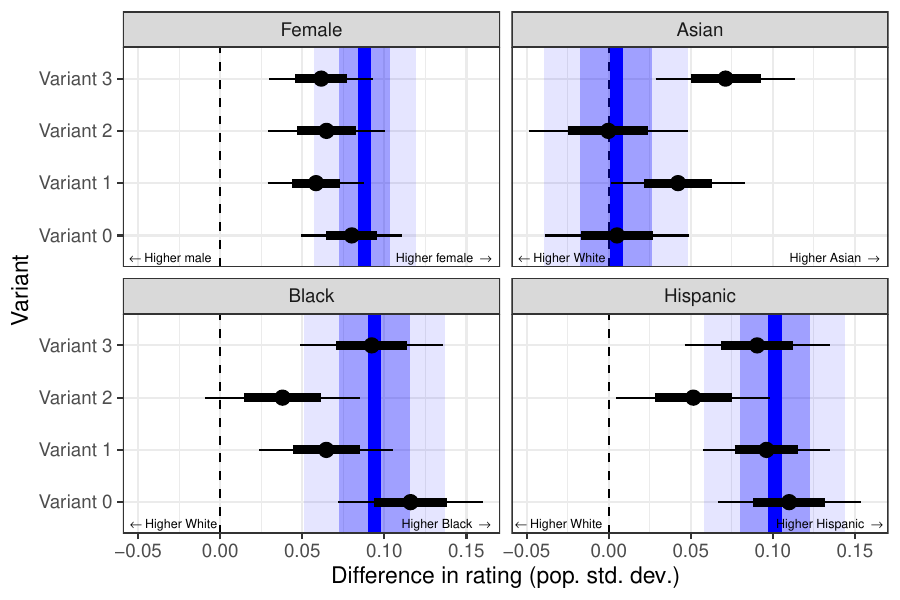}
  \end{center}
  \caption{\emph{%
    Differences in mean model scores across variations in the wording of the
    prompt between synthetic applicants of different races and genders, reported
    in estimated population standard deviations, with 70\% and 95\% confidence
    intervals clustered by real the application dossier used to generate the
    synthetic application. Positive values indicate that the model rates female
    or racial minority applicants higher than male or White applicants on
    average. The blue vertical line represents the estimated effect in the
    original evaluation task, along with 70\% and 95\% confidence intervals.
  }}%
\label{fig:variant}
\end{figure}

To test the sensitivity of our results to variations in the wording of the
prompt, we generate four alternate prompts which are substantially similar to
the prompt used in our primary analysis but differ in the exact choice of words
used to describe the evaluation task. Specifically, using GPT-4, we have the
model translate the original prompt into a foreign language, and then back into
English. The results differ substantially from the original prompt in terms of
word choice (e.g., ``educator'' in place of ``teacher,'' ``suitability'' in
place of ``fit'') but otherwise closely parallel the description of the task. We
try four such variants, finding consistent but modestly variable race- and
gender-effects with the rewritten prompts, as shown in Figure~\ref{fig:variant}.
Here and in the main text, we estimate differences using a linear model where
the outcome is the model’s hiring rating, the covariates are gender and race,
and errors are clustered at the level of the real application that was used to
generate the synthetic application dossier. As before, we standardize the
differences by the estimated population standard deviation.

\begin{figure}[t]
  \begin{center}
    \includegraphics{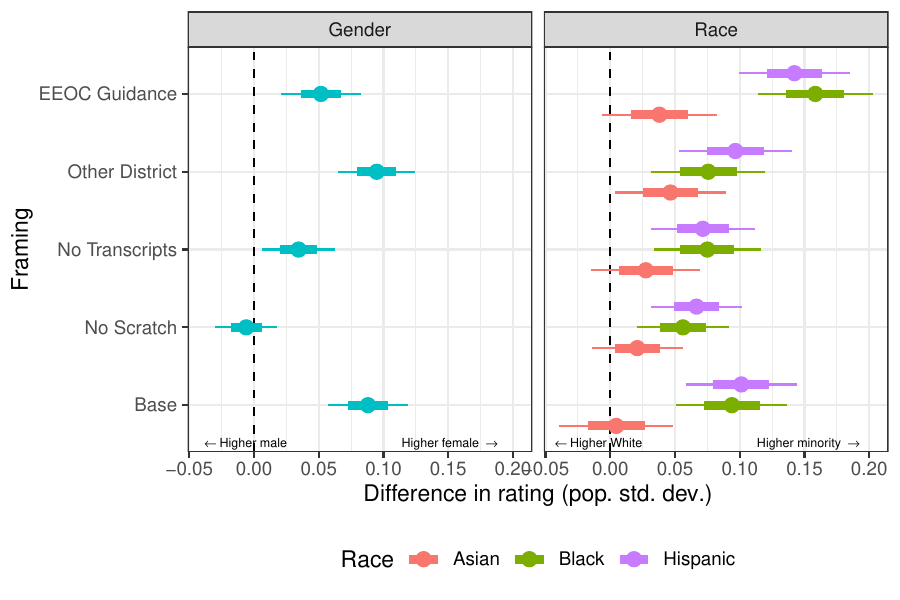}
  \end{center}
  \caption{\emph{%
    Differences in mean model scores across variations in the prompt and context
    between synthetic applicants of different races and genders, reported in
    estimated population standard deviations, with 70\% and 95\% confidence
    intervals clustered by real the application dossier used to generate the
    synthetic application. Positive values indicate that the model rates female
    or racial minority applicants higher than male or White applicants on
    average.
  }}%
\label{fig:framing}
\end{figure}

Here, the blue vertical line represents the estimated race- or gender-effect in
the original evaluation task, along with 70\% and 95\% confidence intervals. As
can be seen, estimated effects vary across conditions, but only slightly beyond
what would be expected on the basis of estimation error alone.

To test the sensitivity to our results to variations in the evaluation task
itself, we also conduct the following variants of our main experiment:
\begin{itemize}
  \item \textbf{``No Scratch'':} We modify the original task, removing the
    elicitation of a summary of the candidate's qualifications.
  \item \textbf{``No Transcripts'':} We elicit hiring recommendations from the
    model as in the original task, but omit the interview transcripts from the
    input to the model.
  \item \textbf{``Other District'':} In applicant responses to interview
    questions and the description of the evaluation task itself, we substitute
    in place of the name of the actual school district to which applicants
    applied (and minor variations of the name that commonly occur in the
    transcripts) the name of an alternate, mostly White school district in West
    Virginia. We also replace the city and state where the actual school
    district is located.
  \item \textbf{``EEOC Guidance'':} We include a brief instruction to adhere to
    the Equal Opportunity Employment Commission’s guidelines on disparate impact
    and disparate treatment in the original task description.
\end{itemize}
The results are shown in Figure~\ref{fig:framing}. As in the case of variations
in the wording of the prompt, the estimated race- and gender-effects vary
modestly, but in many cases within the range we would expect from estimation
error alone.

\subsection*{Impact of Blinding}

Blinding is a natural strategy for mitigating race- and gender-effects in model
evaluations. To understand the impact of blinding, we repeat the adverse impact
analysis in the main text, substituting resumes and transcripts from which name,
pronouns, and other obvious indicators of race and gender have been removed for
the unmodified resume text and transcripts. We find that while model evaluations
do differ, the adverse impact ratio would change relatively little compared to
the adverse impact ratio if the model evaluated unblinded application materials.

\begin{figure}[t]
  \begin{center}
    \includegraphics{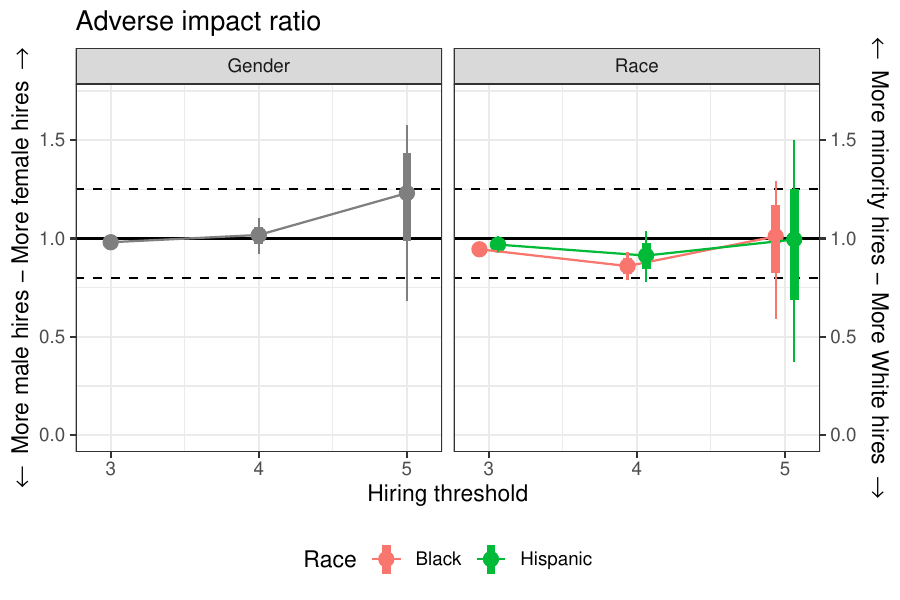}
  \end{center}
  \caption{\emph{%
    Adverse impact ratios for LLM hiring recommendations at different hiring
    thresholds, with pivotal 95\% bootstrapped confidence intervals, when the
    model is presented with redacted application materials. The results are
    substantially similar to those in the main text: we find near parity at the
    lowest threshold, but some evidence of disparities at higher thresholds.
  }}%
\label{fig:adverse-blinded}
\end{figure}

\subsection*{Prediction of Race and Gender from Redacted Materials}

\begin{figure}[t]
  \begin{center}
    \includegraphics{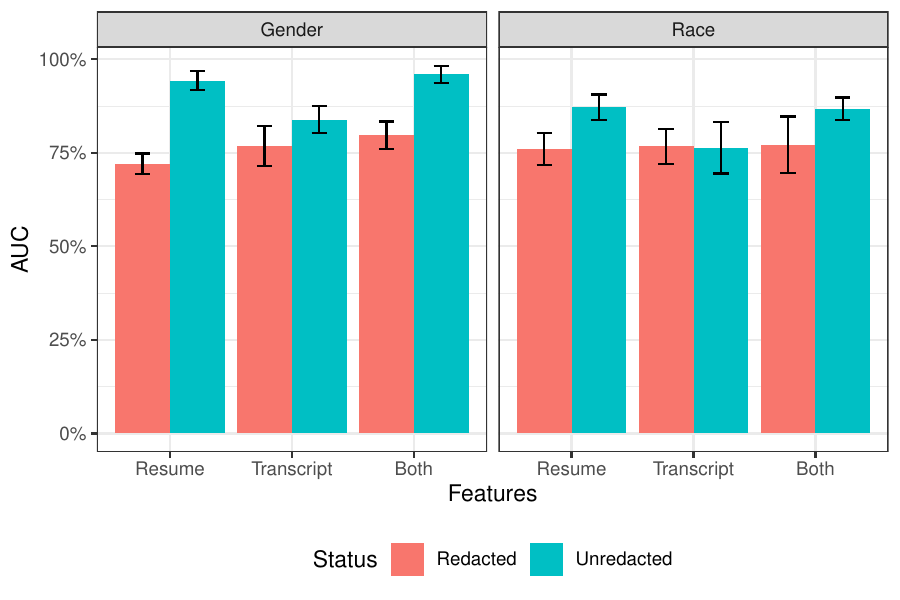}
  \end{center}
  \caption{\emph{%
    Out-of-sample AUC for predicting race and gender from redacted and
    unredacted application materials. The AUC is generally higher for unredacted
    materials, but substantial predictive power remains even when name,
    pronouns, and other obvious indicators of race and gender have been removed.
  }}%
\label{fig:auc}
\end{figure}

Redacting an applicant’s listed name, pronouns, title, and college removes some
information about their race and gender from the application materials; however,
it does not remove all available information. For instance, an applicant may
have held a job that is highly correlated with gender, or speak a dialect of
English strongly associated with a certain race or ethnic group. To test how
much information about race and gender the redacted and unredacted application
materials contain, we embed into 256 dimensions (512 total) both kinds of
application dossiers using OpenAI’s \texttt{text-embedding-small}. We then split
half the data into a training set which we use to train a penalized logistic
regression model to predict applicants’ races and genders using (1) just the
embedding of the resume, (2) just the embedding of the transcripts, and (3) both
embeddings. We then calculate the out-of-sample AUC using 10-fold
cross-validation on the held out testing data. We find that while it is
generally easier to predict race and gender with the unredacted materials, one
can still predict race and gender quite accurately using only redacted
materials; see Figure~\ref{fig:auc}.

\section*{Models}

In our experiments, we used the following model versions through the OpenAI API
and on the AWS Bedrock service, current as of April~1, 2024:
\begin{itemize}
  \item OpenAI
    \begin{itemize}
      \item \textbf{GPT-3.5:} \texttt{gpt-3.5-turbo-0125}
      \item \textbf{GPT-4:} \texttt{gpt-4-0125-preview}
    \end{itemize}
  \item Anthropic
    \begin{itemize}
      \item \textbf{Claude Instant:} \texttt{claude-instant-v1}
      \item \textbf{Claude 2:} \texttt{claude-v2:1}
      \item \textbf{Claude 3 Sonnet:} \texttt{claude-3-sonnet-20240229-v1:0}
      \item \textbf{Claude 3 Haiku:} \texttt{claude-3-haiku-20240307-v1:0}
    \end{itemize}
  \item Mistral
    \begin{itemize}
      \item \textbf{Mistral 7B:} \texttt{mistral-7b-instruct-v0:2}
      \item \textbf{Mixtrax 8x7B:} \texttt{mixtral-8x7b-instruct-v0:1}
    \end{itemize}
\end{itemize}

\end{document}